\documentclass[12pt,twoside,a4paper]{article}
\usepackage[dvips]{epsfig}
\begin{document}
\thispagestyle{empty} 
\title{
\vskip-3cm
{\baselineskip14pt
\centerline{\normalsize DESY 01--134 \hfill ISSN 0418--9833}
\centerline{\normalsize MZ-TH/01--23 \hfill} 
\centerline{\normalsize hep--ph/0109167 \hfill} 
\centerline{\normalsize September 2001 \hfill}} 
\vskip1.5cm
Inclusive $D^{*}$ Production in Photon-Photon Collisions\\
at Next-to-Leading Order QCD 
\author{G.~Kramer$^1$ and H.~Spiesberger$^2$
\vspace{2mm} \\
{\normalsize $^1$ II. Institut f\"ur Theoretische
  Physik\thanks{Supported by Bundesministerium f\"ur Forschung und
    Technologie, Bonn, Germany, under Contract 05~HT9~GA3, and
    by EU Fourth Framework Program {\it Training and Mobility of
    Researchers} through Network {\it Quantum Chromodynamics and
    Deep Structure of Elementary Particles}
    under Contract ERBFMRX--CT98--0194 (DG12 MIHT).}, Universit\"at
  Hamburg,}\\ 
\normalsize{Luruper Chaussee 149, D-22761 Hamburg, Germany} \vspace{2mm}
\\ 
\normalsize{$^2$ Institut f\"ur Physik,
  Johannes-Gutenberg-Universit\"at,}\\ 
\normalsize{Staudinger Weg 7, D-55099 Mainz, Germany} \vspace{2mm} \\
} }

\date{}
\maketitle
\begin{abstract}
\medskip
\noindent
The next-to-leading order cross section for the inclusive production of
charm quarks in $\gamma \gamma $ collisions is calculated as a function
of the transverse momentum $p_T$ and the rapidity $y$ in approaches
using massive or massless charm quarks. For the direct cross section we
derive the massless limit from the massive theory with the result that
this limit differs from the massless version with $\overline{MS}$
factorization by finite corrections. Subtracting or adding these
corrections allows us to compare the two approaches on equal footing. We
establish massless and massive versions with 3 and 4 initial flavours
which are shown to approach the massless approximations very fast with
increasing $p_T$. With these results we calculate the inclusive
$D^{*\pm}$ cross section in $\gamma \gamma $ collisions using realistic
evolved fragmentation functions with appropriate factorization scales
and compare with recent data for $d\sigma/dp_T$ from three LEP
collaborations after single- and double-resolved contributions have been
added.
\end{abstract}

\clearpage

\section{Introduction}

Recently the three LEP collaborations, ALEPH \cite{1}, L3 \cite{2} and
OPAL \cite{3} have presented data for inclusive $D^{*\pm}$ production in
two-photon collisions at $e^+e^-$ center-of-mass energies close to
$\sqrt{S} = 189$ GeV. Besides the total cross section $\sigma$ for
$\gamma + \gamma \rightarrow D^{*} + X$ also the differential cross
sections with respect to the $D^{*}$ transverse momentum,
$d\sigma/dp_T$, and the pseudo-rapidity, $d\sigma/d\eta$, have been
measured.
\\

In $\gamma \gamma$ collisions, where both photons are on-shell, each of
the two photons can behave as either a point-like or a hadron-like (or
resolved) particle. Therefore one distinguishes in such collisions three
production channels called {\em direct} (both particles interact
point-like), {\em single-resolved} (one $\gamma$ is point-like, the
other hadron-like) and {\em double-resolved} (both photons are
hadron-like).  The resolved contributions require the knowledge of
parton densities in the photon, whereas the production via two direct
photons is free of such non-perturbative input. The different channels
mix in higher orders of perturbation theory and thus the distinction
between the direct and resolved contributions becomes scale and scheme
dependent: only the sum of all three contributions is a physical cross
section.
\\

The transverse momentum distribution $d\sigma/dp_T$ for the inclusive
production of $D^{*\pm}$ mesons is characterized by two distinct scales,
the mass $m$ of the charmed quark and the transverse momentum $p_T$ of
the $D^{*}$ or charm-quark. Depending on the ratio $p_T/m$, two
different approaches for next-to-leading order (NLO) calculations in
perturbative QCD have been used for a comparison with the experimental
data \cite{1, 2, 3, 4}. In the so-called {\em massless scheme}, the
charm quark is considered as an active flavour in the photon \cite{5}.
In these calculations, it is assumed that four flavours $q = u$, $d$,
$s$ and $c$ are present in the photon, described by corresponding
distribution functions, and all quarks are taken to be massless. The
charm-quark is also an in-going parton originating from the photon in
the case of the resolved contributions to the cross section and it
fragments into the $D^{*}$ meson similarly as the produced $u$, $d$, $s$
quarks and the gluon $g$. The predictions of this approach are expected
to be valid in the region of large transverse momenta $p_T \gg m$. In
this scheme, calculations for the small $p_T$ region are not reliable.
The cross section diverges in the limit $p_T \rightarrow 0$ and the
total cross section can not be predicted. Following the usage in deep
inelastic charm production \cite{6} we shall refer to this as the {\em
  zero-mass (ZM) 4-flavour scheme} (it corresponds to the "massless
charm scheme"introduced in Ref.\ \cite{7} in connection with charm
production in $\gamma p$ collisions).
\\

The other scheme, in which cross sections for $\gamma \gamma \rightarrow
D^{*}~X$ have been calculated \cite{8}, is the so-called {\em massive
  charm scheme} \cite{7}.  In the massive charm scheme the number of
active massless flavours in the initial state for the resolved
contribution is equal to $n_f=3$ and the charm quark is assumed to be
massive.  The massive $c$ quark appears only in the final state.  In
this scheme the charm mass, $m \gg \Lambda_{QCD}$, acts as a cutoff for
the initial and final state collinear singularities, and sets the scale
for the perturbative calculations.  The cross section factorizes into a
partonic hard scattering cross section multiplied by light quark and
gluon densities in the case of the resolved contributions. In leading
order (LO), the direct production is described by the partonic reaction
$\gamma + \gamma \rightarrow c + \bar{c}$ while the resolved
contributions involve the channels $\gamma + g \rightarrow c + \bar{c}$
(single resolved) and $q + \bar{q} \rightarrow c + \bar{c}$ and $g + g
\rightarrow c + \bar{c}$ (double resolved), where $q$ are the light
(massless) quarks $q = u$, $d$, and $s$. This approach has the advantage
that not only the various distributions, like in rapidity and/or
transverse momentum, can be predicted in the full range of $p_T$ but
also the total cross section.
\\

One might expect that the massive approach is reasonable only in those
kinematical regions where the mass $m$ and any other characteristic
scale like $p_T$ are approximately of the same magnitude and
significantly larger than $\Lambda_{QCD}$. Under these circumstances the
charm mass can be used to set the renormalization scale entering the
quark-gluon coupling $\alpha_s$ as well as the factorization scale
needed to evaluate the quark and gluon densities of the photon. In NLO,
terms $\propto \alpha_s \ln(p_T^2/m^2)$ arise from collinear emission of
the gluon by charmed quarks at large transverse momentum or from almost
collinear branching of photons or gluons into $c\bar{c}$ pairs. These
terms are not expected to affect the total production rates, but they
might spoil the convergence of the perturbation series and cause large
scale dependencies of the NLO result at $p_T \gg m$.\footnote[1]{
  Similar potentially large terms $\propto \alpha_s \ln(Q^2/m^2)$, $Q$
  being the photon virtuality, appear in the calculation of charm
  electro-production cross sections at next-to-leading order \cite{6}.}
In the massive approach the prediction of differential cross sections is
thus limited to a rather small range of $p_T \simeq m$. Nevertheless,
predictions of this approach have been compared to data up to $p_T
\simeq 10$ GeV \cite{8}. 
\\

The proper procedure for $p_T \gg m$ is to absorb potentially large
logarithms into distribution and fragmentation functions where they can
be resummed by virtue of the Altarelli-Parisi equations. To implement
this procedure one needs a charm contribution in the photon parton
distributions (PDF) and a fragmentation function (FF) for the transition
$c \rightarrow D^{*}$ (or any other charmed meson or baryon). Logarithms
$\propto \ln (M^2/m^2)$ defined with the factorization scale $M$ are
absorbed into these distribution and fragmentation functions and
remaining terms $\propto \ln (p_T^2 / M^2)$ are of order $O(1)$ for the
appropriate choice $M \simeq p_T$.  For sufficiently large $p_T$ the
cross sections calculated with this finite charm mass method including
four active flavours must approach the results described earlier with
the zero-mass 4-flavour scheme \cite{5}.  Since $m \neq 0$, it is
reliable also for intermediate $p_T > m$, where most of the experimental
data have been obtained so far, rather than only for $p_T \gg m$. We
shall call this scheme with a massive charm quark, but with terms
proportional to $\alpha_s \ln(M^2/m^2)$ subtracted, the {\it NLO
  4-flavour scheme}.  The scheme described above, where this subtraction
is not performed and only three light active flavours are considered
will be called the {\it NLO 3-flavour scheme}. This latter heavy quark
mass approach is also often referred to as the fixed-flavour number
scheme \cite{6} in connection with heavy quark electro-production. In
the following we shall not use the term "massive charm scheme" any more,
since it is ambiguous. Instead we shall refer either to the NLO
3-flavour or the NLO 4-flavour approach.
\\

The relation of these three approaches for the process $\gamma \gamma
\rightarrow D^{*} X$ has not been investigated in detail yet. In
particular, the relation between the ZM 4-flavour and the NLO 4-flavour
scheme, e.g.\ the question, in which range of $p_T$ values the ZM
4-flavour scheme is a good approximation to the NLO 4-flavour scheme,
has not been studied so far.  Also, it is of interest to know, for which
$p_T$ the NLO 3-flavour and the NLO 4-flavour scheme produce
approximately the same results. So far only differential cross sections
calculated in the ZM 4-flavour and in the NLO 3-flavour scheme have been
compared \cite{9}. It was found that the two approaches differ in the
definitions and relative contributions of the direct and resolved terms,
but essentially agree in their sum. However, in this comparison
\cite{9}, the ZM 4-flavour result was modified in so far as terms
containing $\alpha_s \ln(M^2/m^2)$ have been taken into account not
only at order $O(\alpha_s)$, but resumming them with the
Altarelli-Parisi equations in the so-called perturbative fragmentation
function approach, in which initial conditions for the FF's were taken
from perturbation theory \cite{10}.  This made it difficult to pin down
the finite charm-mass effects present in the NLO 3-flavour scheme.
\\

It is the purpose of this work to fill this gap and present a comparison
of results obtained in the ZM and the $m \neq 0$ 4-flavour schemes. We
will identify terms in the massive theory surviving in the limit $m
\rightarrow 0$ which are not present in the ZM approach where the quark
mass is put to zero from the beginning. These terms describe final-state
interactions and can be interpreted as a perturbative fragmentation
function describing the transition from massless to massive charm
quarks. Only after correcting for this difference by subtracting the
final-state interaction terms from the massive theory (or adding them to
the massless theory) one can expect that both theories approach each
other in the large-$p_T$ limit. This way, the $m = 0$ and the $m \neq 0$
theories can be considered on the same footing and a comparison will be
sensible. In addition we shall compare also to the NLO 3-flavour scheme
in order to find out the difference at small and intermediate $p_T$. We
shall concentrate in this comparison on the direct cross section for
$\gamma \gamma \rightarrow D^{*}~X$, since this is the dominant part. A
similar comparison for the single-resolved and the double-resolved cross
section is left to a later study. The merging of the massive fixed-order
approach and the resummed massless fragmentation approach has been
investigated in Ref.\ \cite{11} for hadron-hadron and photon-hadron
scattering. In principle one can infer from these results the
corrections due to the finite charm mass for the single- and
double-resolved cross sections.  This work, however, as well as the
calculations for $\gamma \gamma$ reactions in Ref.\ \cite{9}, is based
on the perturbative FF's \cite{10} approach, which is too restrictive
for our purpose.
\\

The outline of our work is as follows. In section 2, we shortly describe
the formulae which are used to calculate the cross section for $\gamma +
\gamma \rightarrow c/\bar{c} + X$ with $m \neq 0$ .  We derive from
these cross sections the limit $m \rightarrow 0$ and compare it with the
results of the zero-mass theory which can be found in the literature.
This defines the necessary subtractions in the massive theory so that it
approaches the massless theory in the limit of large $p_T$ . Section 3
contains the numerical results for the comparison of the two theoretical
approaches based on different choices for the scales at which finite
initial and final state terms are subtracted or absorbed into
non-perturbative PDF's of the photon or FF's of the charm quark. In this
section we also present comparisons to the 3-flavour scheme. After
adding single- and double-resolved contributions we compare the results
for the $m \neq 0$ and the $m = 0$ 4-flavour schemes to recent
experimental data from LEP II.  Our conclusions are summarized in
section 5.
\\


\section{Decomposition of the LO and NLO Differential Cross Section}

\subsection{Leading-Order Cross Section}

We first consider the process
\begin{equation}
 \gamma (p_1) + \gamma (p_2) \rightarrow c(p_3) + \bar{c}(p_4) +[g(k)]
\end{equation}
where $p_i$, $i=1$, $2$, $3$, $4$ and $k$ denote the momenta of the two
incoming photons and the outgoing $c$, $\bar{c}$ quarks and a possible
gluon (in square brackets), which is present when we consider the NLO
corrections. Below we will describe the procedure needed to obtain
differential cross sections for $D^*$ production in $\gamma \gamma $
scattering.  We have the following invariants
\begin{equation}
 s = (p_1+p_2)^2, ~~
 t = T-m^2 = (p_1-p_3)^2 - m^2, ~~
 u = U-m^2 = (p_2-p_3)^2 - m^2
\end{equation}
and
\begin{equation}
 s_2 = S_2-m^2 = (p_1+p_2-p_3)^2-m^2 = s+t+u \, .
\end{equation}
It is customary to define the dimensionless variables
\begin{equation}
 v = 1 +\frac{t}{s} ,~~
 w =-\frac{u}{s+t}
\end{equation}
so that
\begin{equation}
 t = -s(1-v), ~~
 u = -svw, ~~
 s_2 = sv(1-w) \, .
\end{equation}
The leading-order cross section is
\begin{eqnarray}
 \frac{d\sigma_{\rm LO}}{dvdw} = c(s) \delta (1-w) 
  \left(\frac{t}{u}+\frac{u}{t} + 4\frac{sm^2}{tu} -
        4\left(\frac{sm^2}{tu}\right)^2
  \right)
\label{sigma_LO}
\end{eqnarray}
where  
\begin{equation}
 c(s) = \frac{2\pi N_C \alpha^2 e_c^4}{s} \, .
\end{equation} 
$N_C$ is the number of quark colours and $e_c$ is the electric charge of
the charm quark, $e_c = 2/3$. From (\ref{sigma_LO}) the finite charm
mass corrections are clearly visible. In the next section we shall show
numerical results explicitly.


\subsection{The Next-to-Leading-Order Cross Section}

The NLO corrections consist of two parts, the virtual corrections to
$\gamma + \gamma \rightarrow c + \bar{c}$ and the gluonic bremsstrahlung
contributions $\gamma + \gamma \rightarrow c + \bar{c} + g$. These NLO
corrections have been calculated by several groups \cite{12, 13, 14, 9}.
Only in Ref.\ \cite{13} explicit formulae for the separate contributions
due to one-loop diagrams and due to bremsstrahlung contributions are
given in a form which allows us to derive the massless limit ($m
\rightarrow 0$).  The results in Ref.\ \cite{13} are subdivided into
three parts, the vertex plus self-energy cross section $d\sigma_{\rm
  VSE}$, the virtual box cross section $d\sigma_{\rm Box}$ and the gluon
bremsstrahlung cross section $d\sigma_{\rm Br}$, so that the NLO
single-inclusive differential cross section $d\sigma/dvdw$ is decomposed
as follows
\begin{equation}
\frac{d\sigma_{\rm NLO}}{dvdw}
= \frac{d\sigma_{\rm VSE}}{dvdw}
+\frac{d\sigma_{\rm Box}}{dvdw} 
+ \frac{d\sigma_{\rm Br}}{dvdw} \, .
\label{sigma_NLO}
\end{equation}
The three parts in (\ref{sigma_NLO}) have according to Ref.\ \cite{13}
the following structure:
\begin{eqnarray}
\frac{d\sigma_{\rm VSE}}{dvdw} 
& = & 
\frac{C(s)}{4}
  \delta(1-w) 
  \Biggl\{2A_1 
    \left( 4 \left[\zeta_2 - {\rm Li}_2 \left(\frac{T}{m^2}\right)
             \right] 
          \left(1+3\frac{m^2}{t}\right) 
    \right. 
\nonumber 
\\
& &
   - \left. 
     \ln\left(\frac{-t}{m^2}\right) 
     \left(8 - 6\frac{t}{T} - \frac{t^2}{T^2}\right)
     - 2 -\frac{t}{T}
     \right)
\nonumber 
\\
& &
  + A_2 \ln\left(\frac{-t}{m^2}\right)
  + A_3 \left[{\rm Li}_2\left(\frac{T}{m^2}\right) - \zeta_2\right] 
  + A_4 + (t \leftrightarrow u)
  \Biggr\}
\label{sigma_VSE}
\end{eqnarray}
and
\begin{eqnarray}
\frac{d\sigma_{\rm Box}}{dvdw} 
& = &
\frac{C_F\alpha_s}{\pi}\frac{d\sigma_{\rm LO}}{dvdw}\frac{2m^2-s}{s\beta}
\left\{ - 2\ln x \ln \beta + 2{\rm Li}_2(-x) - 2{\rm Li}_2(x) - 3\zeta_2
\right\}
\nonumber 
\\
& &
+ \frac{C(s)}{4} \delta (1-w)
  \Biggl\{ 
  -8B_1 \frac{2m^2-s}{s\beta} \ln x \ln\left(\frac{-t}{m^2}\right)
\nonumber 
\\ 
& &
+ 2 \frac{B_2}{\beta}
\left( \ln x \left[4\ln (1+x) - \ln x - 4\ln\left(\frac{-t}{m^2}\right)
             \right]
       + 4{\rm Li}_2(-x) + 2\zeta_2
\right)
\nonumber 
\\
& &
+ 2B_3 \ln^{2}x + 4\frac{B_4}{\beta} \ln x 
+ 4B_5 \ln\left(\frac{-t}{m^2}\right)
+ 8B_6 \ln\left(\frac{T}{m^2}\right)
+ 4B_7 \zeta_2 + 4B_8 
\nonumber
\\
& &
+ (t \leftrightarrow u)
\Biggr\} \, .
\label{sigma_box}
\end{eqnarray}
In (\ref{sigma_box})
\begin{equation}
 \beta = \sqrt{1-4m^2/s} \, , ~~ 
 x = \frac{1-\beta}{1+\beta}
\end{equation} 
and in (\ref{sigma_VSE}) and (\ref{sigma_box}) $\zeta_2=\pi^2/6$ and the
normalisation
\begin{equation}
  C(s) = c(s) \frac{C_F\alpha_s}{2\pi} \, .
\end{equation}
The quantities $A_i$ and $B_i$ are functions of $m^2$, $s$, $t$ and $u$.
They are given in appendix B of Ref.\ \cite{13} and will not be repeated
here.  The contributions (\ref{sigma_VSE}) and (\ref{sigma_box}) contain
also infrared divergent terms proportional to $\epsilon^{-1}$
($2\epsilon = 4-n$) in dimensional regularization with dimension $n$.
They are omitted since they cancel against terms in the bremsstrahlung
cross section. How these terms are distributed in $d\sigma_{\rm VSE}$
and $d\sigma_{\rm Box}$ can be inferred from Ref.\ \cite{13}.
\\

The last term in (\ref{sigma_NLO}) looks more complicated. According to
Ref.\ \cite{13} it can be written in the following form: 
\begin{eqnarray}
\frac{d\sigma_{\rm Br}}{dvdw}
& = & C(s)\Biggl\{\frac{svs_2}{4S_2}
           \left(\frac{s_2(s+u)}{4S_2}e_2
                 + \frac{2S_2}{s_2(s+u)}\ln\frac{S_2}{m^2}e_3
                 + \frac{4S_2}{m^2(s+u)^2}e_4 
           \right.
\nonumber 
\\
& &
            \left.
                 + I_5e_5 + I_8e_8 + I_9e_9 + I_{10}e_{10} 
                 + I_{13}e_{13} + I_{15}e_{15} + I_{16}e_{16} 
                 + (t \leftrightarrow u)
            \right) 
\nonumber 
\\
& &
       + \frac{1}{(1-w)_{+}} \frac{1}{4S_2}
            \Biggl(\tilde{e}_1 
                  + \frac{2S_2}{\tilde{y}} 
                      \ln\frac{T+U-\tilde{y}}{T+U+\tilde{y}}\tilde{e}_6
                  + \frac{4 S_2}{m^2}\tilde{e}_7
                  + s_2 I_{11} \tilde{e}_{11}
\nonumber 
\\
& &
                  + s_2^2 I_{12} \tilde{e}_{12} 
                  + s_2^2 I_{14} \tilde{e}_{14} 
                  + (t \leftrightarrow u)
            \Biggr)
       \Biggr\}
\nonumber 
\\
& &
       + \frac{C_F\alpha_s}{2\pi} \frac{d\sigma_{\rm LO}}{dvdw}
           \frac{1}{s\beta} \Biggl\{ 
                 (2m^2-s)\Biggl(4\ln x \ln \frac{sv}{m^2} + 2\ln x 
\nonumber 
\\
& &
                               - 2\left[
                                 {\rm Li}_2\left(\frac{-4\beta}{(1-\beta)^2}
                                     \right) + \ln^2x
                                  \right]
                         \Biggr)
                         + 2s\beta \left[1 - 2\ln \frac{sv}{m^2}\right]
              \Biggr\} \, .
\label{sigma_Br}
\end{eqnarray}
Here we used $\tilde{y} = \sqrt{\left(t+u\right)^2 - 4m^2 s}$ and the
coefficients $\tilde{e}_1$, $e_2$, $e_3$, $e_4$, $e_5$, $\tilde{e}_6$,
$\tilde{e}_7$, $e_8$, $e_9$, $e_{10}$, $\tilde{e}_{11}$,
$\tilde{e}_{12}$, $e_{13}$, $\tilde{e}_{14}$, $e_{15}$ and $e_{16}$ are
again functions of the invariants $s$, $t$, $u$, $s_2$ and of $m^2$.
$I_5$ to $I_{16}$ are integrals over angles, which have been evaluated
in Ref.\ \cite{13} and are written down in appendix C of this reference.
\\

The cross sections $d\sigma_{\rm VSE}$ and $d\sigma_{\rm Box}$ are
proportional to $\delta(1-w)$. The bremsstrahlung cross section
$d\sigma_{\rm Br}$ contains terms proportional to $\delta(1-w)$ and to
$\left(\frac{1}{1-w}\right)_{+}$. Other contributions are finite for $w
\rightarrow 1$, as long as $m \neq 0$. In the limit $m \rightarrow 0$
they give rise to additional terms proportional to $\delta(1-w)$,
$\left(\frac{1}{1-w}\right)_{+}$, as well as terms proportional to
$\left(\frac{\ln(1-w)}{1-w}\right)_{+}$.
\\

A second approach where the mass of the charm quark is neglected from
the beginning was worked out in Ref.\ \cite{16} and later confirmed in
Ref.\ \cite{15}. With $m = 0$, collinear singularities appear. They are
regularized by dimensional regularization.  In order to understand the
differences of the two approaches, we have evaluated (\ref{sigma_VSE}),
(\ref{sigma_box}) and (\ref{sigma_Br}) in the limit $m \rightarrow 0$.
Special care must be exercised in order to recover all the terms
proportional to $\delta(1-w)$, $\left(\frac{1}{1-w}\right)_{+}$ and
$\left(\frac{\ln(1-w)}{1-w}\right)_{+}$.  We write the result in a form
which has been introduced in the calculation for massless quarks in
Ref.\ \cite{15}. This will allow us to identify the terms which come in
addition to the massless theory in the $\overline{MS}$ factorization
scheme \cite{15}. The LO cross section for $m=0$ has the simple form
\begin{eqnarray}
\lim_{m\rightarrow 0}\frac{d\sigma_{\rm LO}}{dvdw} 
= c(s) \delta(1-w) \tau_0(v)
~~~~~ {\rm with} ~~~~~ 
\tau_0(v) = \frac{v}{1-v} + \frac{1-v}{v} \, .
\label{sigma_LO_massless}
\end{eqnarray}
The decomposition of the NLO cross section in the limit $m \rightarrow
0$ has the form
\begin{eqnarray}
\lim_{m\rightarrow 0}\frac{d\sigma_{\rm NLO}}{dvdw}
& = &
  \left(c_1 + \tilde{\tilde{c}}_1 \ln\frac{m^2}{s}\right) \delta(1-w)
\nonumber 
\\
& &
+ \left(c_2 + \tilde{\tilde{c}}_2 \ln\frac{m^2}{s}
  \right) \left(\frac{1}{1-w}\right)_{+}
+ c_3\left(\frac{\ln (1-w)}{1-w}\right)_{+} 
\nonumber
\\
& &
+ c_5 \ln v + c_6\ln (1-vw) + c_7 \ln (1-v+vw) + c_8\ln (1-v) 
\nonumber 
\\
& &
+ c_9 \ln w + c_{10}\ln (1-w) + c_{11} 
+ \left( \tilde{c}_{11} + \tilde{\tilde{c}}_{11} \right) \ln \frac{m^2}{s} 
\nonumber 
\\
& &
+ c_{12}\frac{\ln (1-v+vw)}{1-w} + c_{13}\frac{\ln w}{1-w}
+ c_{14}\frac{\ln (\frac{1-v}{1-vw})}{1-w} \, .
\label{sigma_massless}
\end{eqnarray}

The coefficients $c_1$, ... $c_{14}$, $\tilde{\tilde{c}}_1$,
$\tilde{\tilde{c}}_2$, $\tilde{c}_{11}$, and $\tilde{\tilde{c}}_{11}$
are either functions of $v$ alone or of $v$ and $w$. We obtained the
following results for them:
\begin{eqnarray}
c_1 & = & C(s)
\left(
     \ln ^2(1-v)\left[\frac{2}{v}+\frac{1}{1-v}-1\right]
     + \ln(1-v)\left[\frac{3}{1-v}-1\right]
\right.
\nonumber 
\\
& &
     + \ln^ 2v\left[\frac{3}{1-v}+\frac{2}{v}-3\right]
     + \ln v\left[-\frac{3}{2(1-v)}+\frac{3}{2v}+2\right]
\nonumber
\\
& &
\left. 
     + \left(4\zeta_2 - \frac{7}{2}\right) \tau_0(v)
\right) 
     + \Delta c_1 \, ,
\label{ci_first}
\end{eqnarray}
where
\begin{eqnarray}
\Delta c_1 = C(s)\left( - 2\ln^2v - 2\ln v + 2
                 \right) \tau_0(v) \, ;
\end{eqnarray}

\begin{eqnarray}
\tilde{\tilde{c}}_1
=
C(s)\left( - 2\ln v - \frac{3}{2}\right) \tau_0(v) \, ; 
\end{eqnarray}

\begin{eqnarray}
c_2
=
C(s)\left(2\ln v - \frac{3}{2}\right) \tau_0(v) + \Delta c_2 \, ,
\end{eqnarray}
where
\begin{eqnarray}
\Delta c_2
=
C(s)\left( - 4\ln v - 2\right) \tau_0(v) \, ;
\end{eqnarray}

\begin{eqnarray}
\tilde{\tilde{c}}_2 = -2 C(s) \tau_0(v) \,;
\end{eqnarray}

\begin{eqnarray}
c_3
=
2C(s) \tau_0(v) + \Delta c_3 \, ,
\end{eqnarray}
where
\begin{eqnarray}
\Delta c_3 
= 
- 4C(s) \tau_0(v) \, ;
\end{eqnarray}

\begin{eqnarray}
c_5
& = &
C(s)\Biggl(\frac{2v(1-v)}{(1-vw)^3} - \frac{2v}{(1-vw)^2} 
          + \frac{v}{(1-v)(1-vw)} + \frac{2v(1-v)}{1-vw} 
\nonumber 
\\
& &
          - \frac{2v}{1-v+vw} - \frac{2}{v} - 4v + \frac{3}{vw}
          - \frac{2}{w} + \frac{2v}{w} + 2w + 4vw
\nonumber
\\
& &
          - \frac{2w}{1-v} + \frac{2w}{v}
    \Biggr) 
+ \Delta c_5 \, ,
\end{eqnarray}
where
\begin{eqnarray}
\Delta c_5
=
C(s)\left(\frac{2v}{1-v} + \frac{2v^2w}{1-v} + \frac{4}{w}
          - \frac{2v}{w} - \frac{4}{vw} + \frac{4v}{1-v+vw}
    \right) \, ;
\end{eqnarray}

\begin{eqnarray}
c_6 
=
2C(s)\left(\frac{1}{v(1-v)w}\left(2v^3 - 2 - 3v^2 + 5v\right)
           + \frac{1}{1-v}\left(v^2w-2v^2\right)
     \right) \, ;
\end{eqnarray}

\begin{eqnarray}
c_7 
= 
-C(s)\left(\frac{4v}{1-v+vw} + \frac{2}{vw}\right) \, ;
\end{eqnarray}

\begin{eqnarray}
c_8 
= 
\frac{2C(s)}{vw}\left(1+(1-v)^2\right) \, ;
\end{eqnarray}

\begin{eqnarray}
c_9 
= 
-\frac{2C(s)}{v(1-v)w}\left(v^3w^2 + v^2w - (1-v)\right) \, ;
\end{eqnarray}

\begin{eqnarray}
c_{10} 
= 
c_5 + \frac{C(s)}{w(1-v)}\left(4v^2w - 2v^2w^2 - 2v^2 - 2\right) \, ;
\end{eqnarray}

\begin{eqnarray}
c_{11}
& = &
C(s)\Biggl(\frac{2v}{1-v+vw} - \frac{8v(1-v)}{(1-vw)^3} 
         + \frac{7v}{(1-vw)^2}
\nonumber 
\\
& &
         - \frac{v}{(1-v)(1-vw)}\left(2-7v+3v^2\right) + 4v - 2 
         + \frac{7}{v} + \frac{1}{1-v} 
\nonumber 
\\
& &
         - w\left(v-2+\frac{8}{v}+\frac{2}{1-v}\right)
         + \frac{2v}{w} - \frac{1}{w} + \frac{1}{2vw}
    \Biggr)
+ \Delta c_{11} \, ,
\end{eqnarray}
where
\begin{eqnarray}
\Delta c_{11}
& = &
C(s)\left(\frac{2v}{1-v+vw} + \frac{1}{1-v} + \frac{w}{1-v} - w(1+v)
    \right.
\nonumber 
\\
& &
    \left.
          + \frac{2}{w} - \frac{v}{w} - \frac{2}{vw} - 1
    \right) \, ;
\end{eqnarray}

\begin{eqnarray}
\tilde{c}_{11}
& = &
C(s)\left( - \frac{2v(1-v)}{(1-vw)^3} + \frac{2v}{(1-vw)^2}
           - \frac{v}{(1-v)(1-vw)} - \frac{2v(1-v)}{1-vw} 
    \right.
\nonumber 
\\
& &
    \left.
           - \frac{v}{1-v} + 4v + \frac{2}{v} - \frac{1}{vw}
           - \frac{v}{w} - 2vw + \frac{v^2w}{1-v} - \frac{2w}{v}
    \right) \, ;
\label{tildec11}
\end{eqnarray}

\begin{eqnarray}
\tilde{\tilde{c}}_{11}
=
C(s)\left(\frac{2v}{1-v+vw} + \frac{v}{1-v}
          -\frac{1+(1-v)^2}{vw} + \frac{v^2w}{1-v}
    \right) \, ;
\label{tildetildec11}
\end{eqnarray}

\begin{eqnarray}
c_{12}
= 
-\frac{2C(s)}{v(1-v)}\left(1+2v(1-v)\right) \, ;
\end{eqnarray}

\begin{eqnarray}
c_{13}
=
2C(s)\frac{1+v^2}{v(1-v)} \, ;
\end{eqnarray}

\begin{eqnarray}
c_{14}
=
2C(s)\frac{1+(1-v)^2}{v(1-v)}
\label{ci_last}
\end{eqnarray}

By comparing with appendix A of Ref.\ \cite{15} we obtain agreement if
we set in (\ref{sigma_massless}) the factorization scales $M_I^2 = M_F^2
= m^2$ and put $\Delta c_1 = \Delta c_2 = \Delta c_3 = \Delta c_5 =
\Delta c_{11} = 0$ in the formulae above. This means, that the $m
\rightarrow 0$ limit of the NLO correction as derived in Ref.\ \cite{13}
agrees with Ref.\ \cite{15} with initial factorization scale $M_I^2 =
m^2$ and final state factorization scale $M_F^2 = m^2$ and all $\Delta
c_i$ terms put to zero. Denoting the ZM NLO result of Ref.\ \cite{15}
with $d\sigma_{\rm ZM}/dvdw$, we can write
\begin{equation}
\lim_{m\rightarrow 0} \frac{d\sigma_{\rm NLO}}{dvdw} 
= \frac{d\sigma_{\rm ZM}}{dvdw}\left(M_I = M_F = m\right) 
+ \frac{d\sigma_{\rm FSI}}{dvdw}
\end{equation} 
with
\begin{equation}
\frac{d\sigma_{\rm FSI}}{dvdw}
= \Delta c_1 \delta(1-w) + \Delta c_2 \left(\frac{1}{(1-w)}\right)_+ \
+ \Delta c_3 \left(\frac{\ln(1-w)}{1-w}\right)_+ 
+ \Delta c_5 \ln v + \Delta c_{11} \, .
\end{equation} 
In Ref.\ \cite{15} terms containing initial and final state
singularities have been subtracted using different factorization scales
$M_I$ and $M_F$, respectively. Therefore, a separation of terms
proportional to $\ln (M_I^2/s)$ (coefficient $\tilde{c}_{11}$) and $\ln
(M_F^2/s)$ (coefficient $\tilde{\tilde{c}}_{11}$) was possible. Starting
from a calculation with $m \neq 0$, this separation can, of course, not
be obtained -- both terms appear with the logarithm $\ln (m^2/s)$. In
our formulas (\ref{tildec11}, \ref{tildetildec11}) we have split this
sum in two terms so that they agree with $\tilde{c}_{11}$ and
$\tilde{\tilde{c}}_{11}$ in Ref.\ \cite{15}.  In Ref.\ \cite{15} the
$\overline{MS}$ factorization scheme is defined with the spin averaging
for the incoming photons taken to be $1/2(1-\epsilon)$ and not equal to
$1/2$ as it is sometimes done.  From our calculation we conclude that
the zero-mass limit of the massive theory automatically yields the
result for the $\overline{MS}$ scheme with $n$-dimensional photon spin
averaging. By comparing with Ref.\ \cite{15} we note that
$\tilde{\tilde{c}}_1$ and $\tilde{\tilde{c}}_2$ originate from the
final-state logarithmic terms.  The initial-state logarithms have no
factors proportional to $\delta(1-w)$ or $1/(1-w)_{+}$ and appear only
in $\tilde{c}_{11}$, i.e.\ in the non-singular contribution for $w
\rightarrow 1$. There are also no terms proportional to $\ln(\mu^2/s)$,
where $\mu$ is the renormalization scale, since $\alpha_s$ is not
renormalized in first order.

Our derivation shows that the massive theory for $\gamma + \gamma
\rightarrow c/\bar{c} +X$ in the limit $m \rightarrow 0$ approaches the
massless theory in $\overline{MS}$ factorization with scales $M_I^2 =
M_F^2 = m^2$ only if the terms $\Delta c_1$, $\Delta c_2$, ..., $\Delta
c_{11}$ are removed. This result is to be expected since the
regularization of collinear singularities with a mass parameter $m$ does
not necessarily give identical results as with dimensional
regularization and $m = 0$ from the beginning. The difference is a sum
of finite terms, which have first been derived by calculating the cross
section for $e^+ + e^- \rightarrow c/\bar{c} + X$ in the two theories
\cite{10}. In Ref.\ \cite{10} it was shown that the additional finite
terms can be interpreted as a partonic fragmentation function $d_c^c(x,
\mu_0)$ for the transition from massless to massive charm quarks, $c(m =
0) \rightarrow c(m \neq 0)$. The massless theory with $\overline{MS}$
factorization has to be folded with $d_c^c(x, \mu_0)$ in order to obtain
the cross section in the zero-mass limit of the massive theory. This
partonic or perturbative FF in $O(\alpha_s)$ is
\begin{equation}
d_c^c(x,\mu_0) 
= 
\frac{C_F\alpha_s}{2\pi}
\left[\frac{1+x^2}{1-x}
     \left(\ln \frac{\mu_0^2}{m^2} - 2\ln (1-x) - 2
     \right)
\right]_{+} \, .
\label{dcc}
\end{equation}
One can show that the extra terms $d\sigma_{\rm FSI}/dvdw$ can be
recovered in the framework of the massless theory in Ref.\ \cite{15} by
convoluting $d_c^c(x,\mu_0 = m)$ as a FF of the charm quark with the
massless LO cross section\footnote{The explicit form for this
  convolution and prescriptions for the rescaling of the kinematic
  variables $v$ and $w$ can be found in \cite{16}.}:
\begin{equation}
\frac{d\sigma_{\rm FSI}(p_T)}{dvdw} = 
\lim_{m\rightarrow 0} \frac{d\sigma_{\rm LO}(\hat{p}_T = p_T/x)}{dvdw} 
\otimes d_c^c\left(x, \mu_0 = m\right) \, .
\end{equation}
This means that the $\Delta c_i$ terms represent final-state
interactions which survive in the massive theory even in the limit $m
\rightarrow 0$. Therefore we have labeled their sum with the subscript
``FSI''. They are non-singular, i.e.\ not proportional to $\ln(m^2/s)$.
\\

In Ref.\ \cite{9} the expression (\ref{dcc}) was used as a starting
condition at the scale $\mu_0 = m$ for the calculation of perturbative
FF's at an arbitrary scale via the usual Altarelli-Parisi evolution
equations in the massless approach. Our derivation shows that we obtain
the same finite final-state interaction terms as in $e^+e^- \rightarrow
c/\bar{c}~X$.  One expects that they appear in any process for which
cross sections in the limit of the massive theory and the massless
theory with common factorization are compared in NLO. We remark that
such finite terms do not appear in connection with initial state
factorization, since the distribution of quarks in the photon has
identical non-singular terms for the two cases where one considers the
massless limit of the massive theory and the theory with massless
quarks. 
\\

The final result for the massless limit of the massive theory, as given
by (\ref{sigma_massless}) with the coefficients written down in
(\ref{ci_first}) to (\ref{ci_last}), can be used in two ways. On the one
hand, we can compare the cross section of the original massive theory
with the massless theory by calculating the cross section from the full
formula (\ref{sigma_NLO}) together with (\ref{sigma_VSE}),
(\ref{sigma_box}) and (\ref{sigma_Br}) with the limit result in
(\ref{sigma_massless}). In this way we can establish how the additional
terms proportional to $m^2$ and possibly $m^2\ln m^2$, i.e.\ all terms
which vanish for $m^2 \rightarrow 0$, depend on the kinematic variables.
In particular, we can find out, in which kinematical region, the
massless limit in (\ref{sigma_massless}) is a good approximation to the
massive theory.  Compared to the massless cross section as derived in
Ref.\ \cite{15} the cross section in (\ref{sigma_massless}) includes the
finite final-state interaction terms $\Delta c_1$, $\Delta c_2$, ...,
$\Delta c_{11}$, and is computed for the factorization scales $M_I^2 =
M_F^2 = m^2$.  On the other hand, it is clear that the massless
approximation (\ref{sigma_massless}) is relevant only for $p_T^2 \gg
m^2$. In this case $p_T^2$ is the large scale, so that $m^2$ is not a
good choice for the factorization scale. The appropriate choice is
$M_I^2 = M_F^2 = \xi \left(m^2 + p_T^2\right)$ with $\xi = O(1)$.  Thus
it makes much more sense to compare the two theories, massive and
massless, for this choice of the factorization scales.  Then the large
logarithms $\propto \ln(s/m^2)$ are removed and absorbed in the PDF of
the photon or in the FF of the $D^{*}$ meson. These PDF's and FF's are
in NLO usually constructed in the $\overline{MS}$ factorization scheme
which we shall assume in the following. In order to establish the effect
of terms proportional to $m^2$ in this more realistic case, we subtract
the finite final-state interaction terms $\Delta c_1$, $\Delta c_2$,
..., $\Delta c_{11}$ from the massive theory and change the
factorization scale in the massive theory to $M_I^2 = M_F^2 = \xi
\left(m^2 + p_T^2\right)$ by additionally subtracting
\begin{eqnarray}
\left(\frac{d\sigma}{dvdw}\right)_{\rm subtr}
= 
  \tilde{\tilde{c}}_1 \ln\frac{m^2}{M_F^2} \delta(1-w) 
\nonumber 
\\
+ \tilde{\tilde{c}}_2 \ln \frac{m^2}{M_F^2} \frac{1}{(1-w)_{+}} 
+ \tilde{c}_{11} \ln \frac{m^2}{M_I^2}  
+ \tilde{\tilde{c}}_{11} \ln \frac{m^2}{M_F^2}
\label{subtraction}
\end{eqnarray}

from the massive cross section above or equally from the massless cross
section in (\ref{sigma_massless}), which then is identical to the result
in Ref.\ \cite{15}. Of course, the difference of the massive and the
massless approximation will be the same as in the first case with $m^2$
as factorization scale, only the relative amount will change. It is
clear that the two cross sections, massive or massless, for $M_I^2 =
M_F^2 = \xi \left(m^2 + p_T^2\right)$ must be supplemented with a
realistic choice for the $c \rightarrow D^{*}$ FF and with the
additional single- and double-resolved cross sections having a charm
contribution in the photon PDF. We remark that the latter approach makes
sense only in the NLO 4-flavour scheme whereas the first approach with
$M_I^2 = M_F^2 = m^2$ is meaningful only in the NLO 3-flavour scheme.
\\


\section{Numerical Results}

We start with presenting results for the $p_T$ distribution of the cross
section for
\begin{equation}
 e^+ (p_+) + e^- (p_-) \rightarrow c(p_3) + X \, . 
\end{equation}
For this, we have to fold the cross section given in the previous
section with the quasi-real photon spectrum given in the
Weizs\"acker-Williams approximation by
\begin{eqnarray}
f_{\gamma}(x) 
& = & 
\frac{\alpha}{2\pi} \left\{ 
\frac{1 + (1-x)^2}{x} 
\ln \frac{E^2 \theta_c^2 (1-x)^2 + m_e^2 x^2}{m_e^2 x^2}
                    \right.
\nonumber
\\
& &
                    \left.
+ 2(1-x) \left[ \frac{m_e^2 x}{E^2 \theta_c^2 (1-x)^2 + m_e^2 x^2}
- \frac{1}{x} \right] 
                    \right\} \, .
\label{WWA}
\end{eqnarray}
Here $E$ is the $e^+$ ($e^-$) beam energy, $\theta_c$ the maximal angle
under which the outgoing electrons (positrons) are tagged and $x =
E_{\gamma} / E$ the fraction of the beam energy entering the 
$\gamma \gamma$ cross section. Fixing the transverse momentum $p_T$ of
the charm quark and its rapidity $y$, one can perform the integration over the
energy fractions of the two photons $x_i = E_i / E$, $i = 1$, 2, with
the result:
\begin{equation}
\frac{d\sigma}{dp_T dy} = \frac{2p_T}{S} 
\int_{VW}^V \frac{dv}{v(1-v)}
\int_{VW/v}^1 \frac{dw}{w} 
f_{\gamma}\left(x_1 = \frac{VW}{vw}\right)
f_{\gamma}\left(x_2 = \frac{1-V}{1-v}\right) 
\left. \frac{d\sigma}{dvdw} \right|_{s=x_1 x_2 S} \, ,
\label{dspty}
\end{equation}
where
\begin{equation}
V = 1 - \sqrt{\frac{p_T^2 + m^2}{S}} e^{-y} \, , ~~~~
W = \frac{1}{V} \sqrt{\frac{p_T^2 + m^2}{S}} e^{y} \, , ~~~~
S = 4 E^2 \, .
\end{equation}
In the calculation with massive charm, the charm mass is kept non-zero
everywhere, also in the definition of kinematic variables and phase
space limits. In particular, $y$ is the rapidity defined by $y =
\frac{1}{2} \ln \frac{E_c + p_{L,c}}{E_c - p_{L,c}}$.  For the
comparison between various versions of the NLO cross section and their
massless limits we consider the $p_T$ distribution $d\sigma/dydp_T$ as a
function of $p_T$ for $y=0$. Since the cross section as a function of
$y$ for fixed $p_T$ is maximal at $y=0$ we expect a similar behavior for
the cross section $d\sigma/dp_T$ after integration over $y$.
\\

In all the following comparisons the charm mass has the value $m = 1.5$
GeV. The total energy is $\sqrt{S} = 2E = 189$ GeV and the maximal angle
in (\ref{WWA}) is $\theta_c = 0.033$ in accordance with the choice in
the OPAL experiment \cite{3}.  The normalization of the cross section,
as given by (\ref{dspty}) and (\ref{sigma_LO}) (or
(\ref{sigma_LO_massless})) in the LO case is changed by multiplication
with the branching ratio $BR(c \rightarrow D^{*}) = 0.235$. This value
for $BR$ is the average of the five experimental measurements collected
and discussed in Ref.\ \cite{17} ($BR= 0.235 \pm 0.007(\pm 0.007)$). A
more recent value for this branching ratio based on measurements of the
four LEP and the SLD experiments is $BR = 0.241 \pm 0.008$ \cite{17a}.
The calculated cross sections are for the sum of $c$ and $\bar{c}$
production.  Including the branching ratio for $c \rightarrow D^{*}$ in
the cross section will make it easier to compare the results for the
cross sections without explicit FF for $c \rightarrow D^{*}$ with the
cross sections with FF included at the end of this section. We should
stress that the figures of this subsection do not include the resolved
contributions so that a comparison with experimental data is not yet
reasonable.
\\

\begin{figure}[th] 
\unitlength 1mm
\begin{picture}(160,85)
\put(0,-1){\epsfig{file=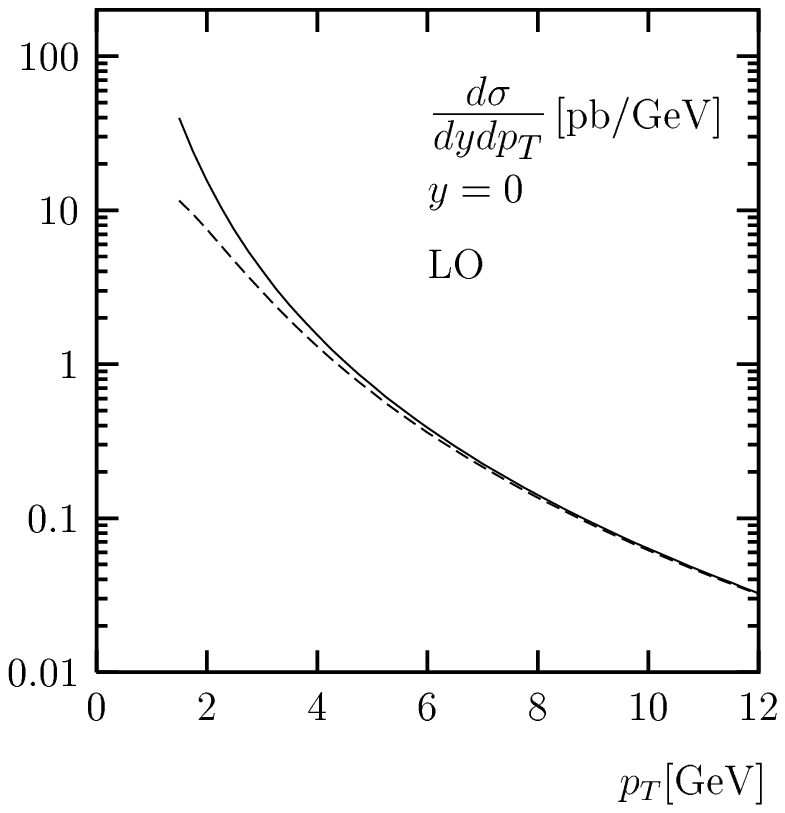,width=8cm}}
\put(40,-1){(a)}
\put(80,-1){\epsfig{file=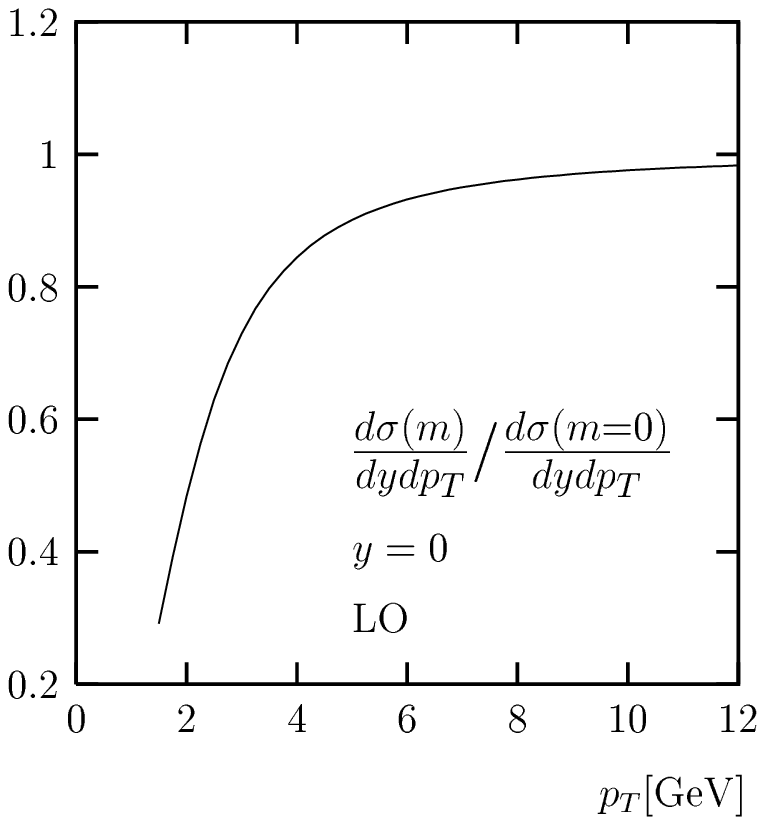,width=8cm}}
\put(120,-1){(b)}
\end{picture}
\caption{(a) $p_T$ distribution $d\sigma/dydp_T$ for $e^+e^-
  \rightarrow \gamma \gamma \rightarrow c/\bar{c} + X \rightarrow
  D^{\ast\pm} + X$ (direct contribution, $BR(c \rightarrow D^{\ast\pm})
  = 0.235$) at LO for $y=0$ in the massless (full line) and the massive
  calculation (dashed line) with $m=1.5$ GeV; (b) ratio of the massive
  and massless calculations.}
\label{fig1}
\end{figure}

We start with the LO cross section calculated from (\ref{dspty}) with
$d\sigma/dvdw$ given in (\ref{sigma_LO}) and its limit for $m=0$ in
(\ref{sigma_LO_massless}). The result for the two cross sections
$d\sigma/dydp_T$ for $m \neq 0$ and $m=0$, respectively, is shown in
Fig.\ \ref{fig1}a.  For large $p_T$ the two cross sections approach each
other. Their ratio, $d\sigma(m) / dydp_T / d\sigma(m=0) / dydp_T$, i.e.\ 
the massive over the massless cross section, as a function of $p_T$, is
plotted in Fig.\ \ref{fig1}b.  As is seen in this figure, the deviation
of the massless from the massive cross section is appreciable at small
$p_T$. The massless cross section increases strongly for $p_T
\rightarrow 0$, whereas the massive cross section increases only
moderately with decreasing $p_T$, making the ratio smaller than one. For
$p_T > 5$ GeV the massive cross section differs by less than $10\,\%$
from the massless cross section. In Fig.\ \ref{fig1}b the ratio crosses
the line $0.9$ at $p_T = 4.9$ GeV.
\\

\begin{figure}[th] 
\unitlength 1mm
\begin{picture}(160,85)
\put(0,-1){\epsfig{file=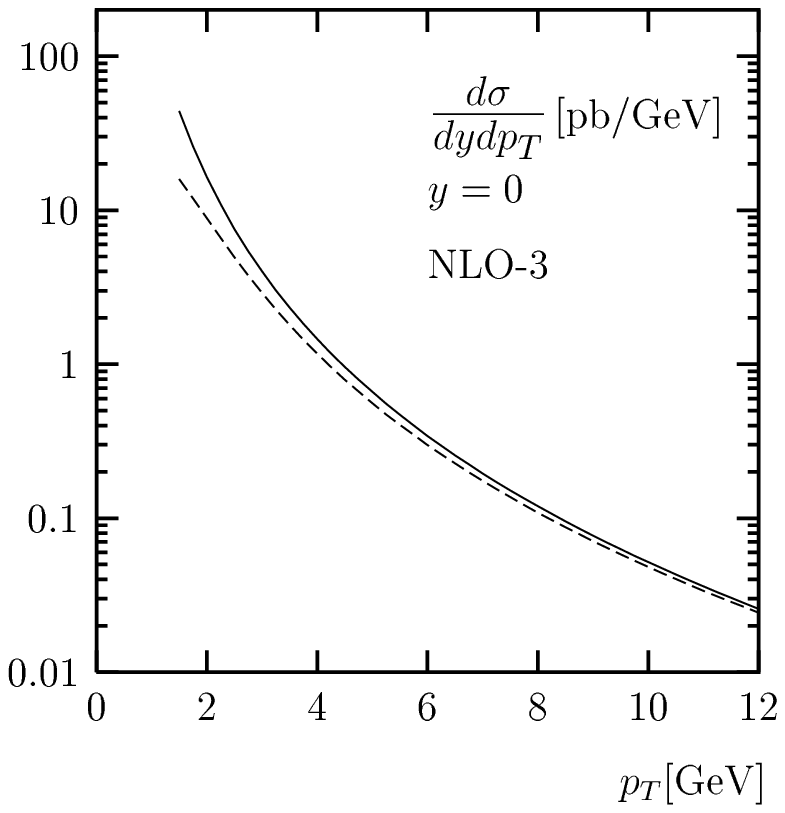,width=8cm}}
\put(40,-1){(a)}
\put(80,-1){\epsfig{file=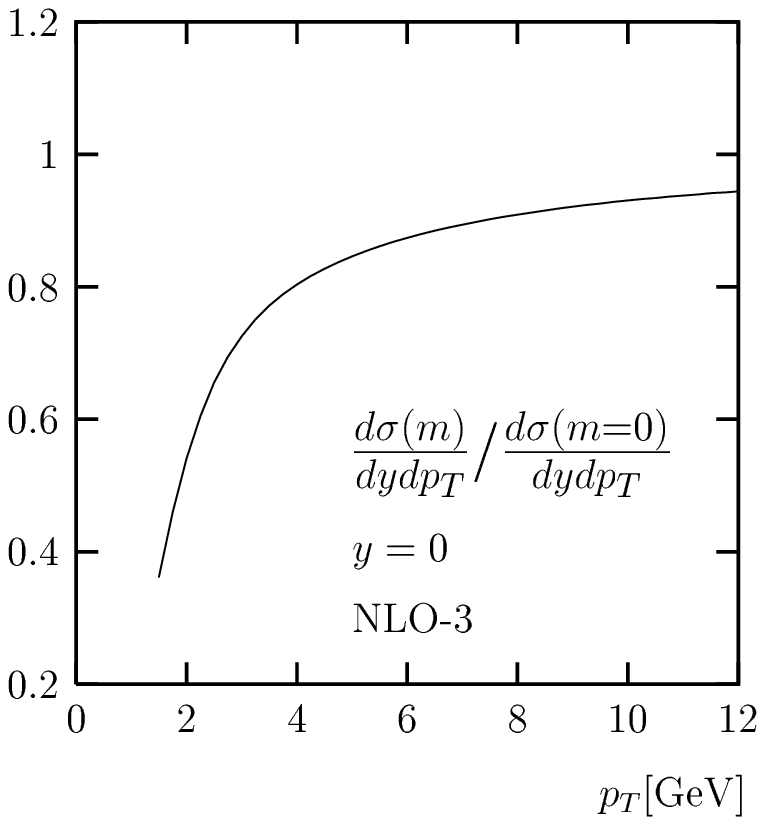,width=8cm}}
\put(120,-1){(b)}
\end{picture}
\caption{(a) $p_T$ distribution $d\sigma/dydp_T$ in the NLO 3-flavour
  scheme (FSI-coefficients $\Delta c_i$ added in the massless
  calculation, direct contribution including $BR(c \rightarrow
  D^{\ast\pm}) = 0.235$) for $y=0$ in the massless (full line) and the
  massive calculation (dashed line) for $M_I = M_F = m$; (b) ratio of
  the massive and massless calculations.}
\label{fig2}
\end{figure}

\begin{figure}[th] 
\unitlength 1mm
\begin{picture}(160,85)
\put(0,-1){\epsfig{file=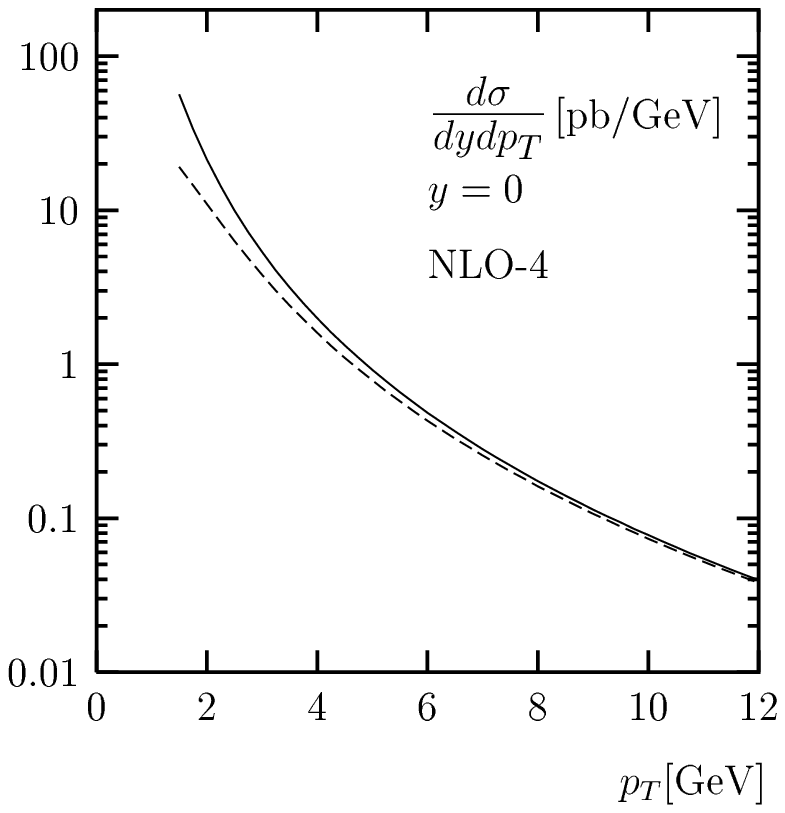,width=8cm}}
\put(40,-1){(a)}
\put(80,-1){\epsfig{file=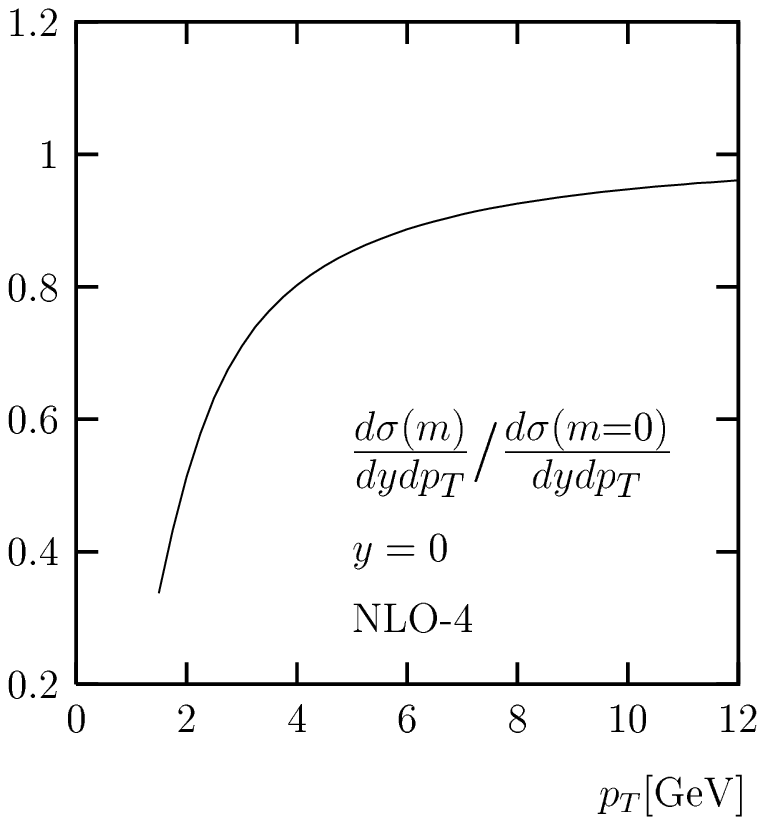,width=8cm}}
\put(120,-1){(b)}
\end{picture}
\caption{(a) $p_T$ distribution $d\sigma/dydp_T$ in the NLO 4-flavour
  scheme (FSI-coefficients $\Delta c_i = 0$ in the massless calculation
  not resummed and subtracted in the massive calculation, direct
  contribution including $BR(c \rightarrow D^{\ast\pm}) = 0.235$) for
  $y=0$ in the massless (full line) and the massive calculation (dashed
  line) for $M_I = M_F = \sqrt{p_T^2 + m^2}$; (b) ratio of the massive
  and massless calculations.}
\label{fig3}
\end{figure}

In the next two figures, Fig.\ \ref{fig2} and Fig.\ \ref{fig3}, we show
the same comparison including the full NLO corrections. The value for
$\alpha_s$ is calculated from the two-loop formula with
$\Lambda_{(N_f=4)} = 328$ MeV which corresponds to $\alpha_s(m_Z) =
0.1181$ in accordance with Ref.\ \cite{18}. The renormalization scale is
always $\mu _R=\sqrt{p_T^2+m^2}$ in all the results of this and the
following section.
\\

In Fig.\ \ref{fig2}a the cross section $d\sigma/dydp_T$ at $y=0$ is
shown in the NLO 3-flavour scheme. For $m \neq 0$ this is the full cross
section with $d\sigma/dvdw$ in (\ref{dspty}) calculated from
(\ref{sigma_NLO}-\ref{sigma_Br}). In the limit $m \rightarrow 0$ we have
added the final-state-interaction coefficients $\Delta c_i$ and changed
the factorization scales to $M_I=M_F=m$. Only with these changes, we can
expect that massive and massless cross sections approach each other for
large $p_T$. This is indeed the case as seen in Fig.\ \ref{fig2}a,b.  If
we compare with the LO cross section in Fig.\ \ref{fig1}a we notice that
the NLO cross section is smaller than the LO one, in particular for
increasing $p_T$, i.e.\ $p_T \geq 3$ GeV. This reduction of the NLO
3-flavour cross section is due to the choice of the factorization scale
$M_I=M_F=m$, which, of course, is the appropriate scale in the case of
fixed three flavours. This generates negative terms in the NLO
corrections which increase in absolute value with increasing $p_T$.
These large correction terms proportional to $\ln (m^2/s)$ (see
(\ref{sigma_massless})) make the perturbative cross section more and
more unreliable, the larger $p_T$ is, and call for a subtraction of
these terms by choosing a factorization scale determined by the large
scale $p_T$. The subtraction terms originate predominantly from the
collinear singularities in the final state and, only to a smaller
extent, from the collinear singularities of the initial state (see
(\ref{subtraction}) for the separation of the two parts). We remark that
we used the same definition of $\alpha_s$ in both schemes, i.e.\ we took
$N_f=4$ and the same value for $\Lambda_{(N_f=4)}$ in the formula for
$\alpha_s$, although, for consistency with the choice of three flavours,
we should have taken $N_f=3$ and the corresponding value for
$\Lambda_{(N_f=3)}$.  This prescription will make it easier to compare
with the results in the NLO 4-flavour scheme. The correct choice with
$N_f = 3$ in $\alpha_s$ would have decreased the cross section shown in
Fig.\ \ref{fig2}a by a few per cent.  The ratio of the massive to the
massless cross section in the NLO 3-flavour scheme is shown in Fig.\ 
\ref{fig2}b as a function of $p_T$. The approach towards unity with
increasing $p_T$ is slower than for the LO ratio in Fig.\ \ref{fig1}b.
This is presumably caused by additional terms $\propto m^2\ln m^2$ which
are not present in LO. Now the ratio is $ \geq 0.9$ for $p_T \geq 7.3$
GeV, i.e.\ in the NLO 3-flavour scheme, larger values of $p_T$ are
required if the massless cross section should be a reasonable
approximation.  Towards smaller $p_T$ the massless cross section
overestimates the full massive cross section quite strongly as was
already noticed in Ref.\ \cite{8}.  At $p_T = 2$ GeV this overestimation
amounts to approximately $85\,\%$.
\\

The cross sections in the NLO 4-flavour scheme are plotted in Fig.\ 
\ref{fig3}a, again for $m \neq 0$ and $m=0$. For $m \neq 0$ the cross
section differs from the one shown in Fig.\ \ref{fig2}a by subtraction
of the FSI-coefficients and by changing the factorization scales to $M_I
= M_F = \sqrt{p_T^2+m^2}$, i.e.\ we subtracted from the massive cross
section (\ref{sigma_NLO}-\ref{sigma_Br}) the terms $\Delta c_i$ and the
logarithms (\ref{subtraction}).  The massless cross section is obtained
from (\ref{sigma_massless}, \ref{ci_first}-\ref{ci_last}) with $\Delta
c_i = 0$, i.e.\ it is identical to the results of Ref.\ \cite{15}. We
observe that the NLO 4-flavour cross section is larger than the NLO
3-flavour cross section in Fig.\ \ref{fig2}a. About half of this
difference is due to the enlarged factorization scale; the other half is
due to the subtraction of the FSI-coefficients $\Delta c_i$. The
increase is most effective at the larger $p_T$ values.  Without
additional fragmentation effects due to the $c \rightarrow D^{*}$
transition the NLO 4-flavour scheme would thus lead to larger cross
sections than the NLO 3-flavour scheme.  The ratio of the massive and
the massless cross section in the NLO 4-flavour scheme is presented in
Fig.\ \ref{fig3}b.  It is clear that the difference between these two
cross sections is equal to the one in the NLO 3-flavour scheme shown in
Fig.\ \ref{fig2}a, but the ratio changes slightly.  Now, the ratio is
larger than $0.9$ at $p_T=6.5$ GeV, i.e.\ at somewhat smaller $p_T$ than
in the NLO 3-flavour scheme (Fig.\ \ref{fig2}b).
\\

\begin{figure}[th] 
\unitlength 1mm
\begin{picture}(160,85)
\put(0,-1){\epsfig{file=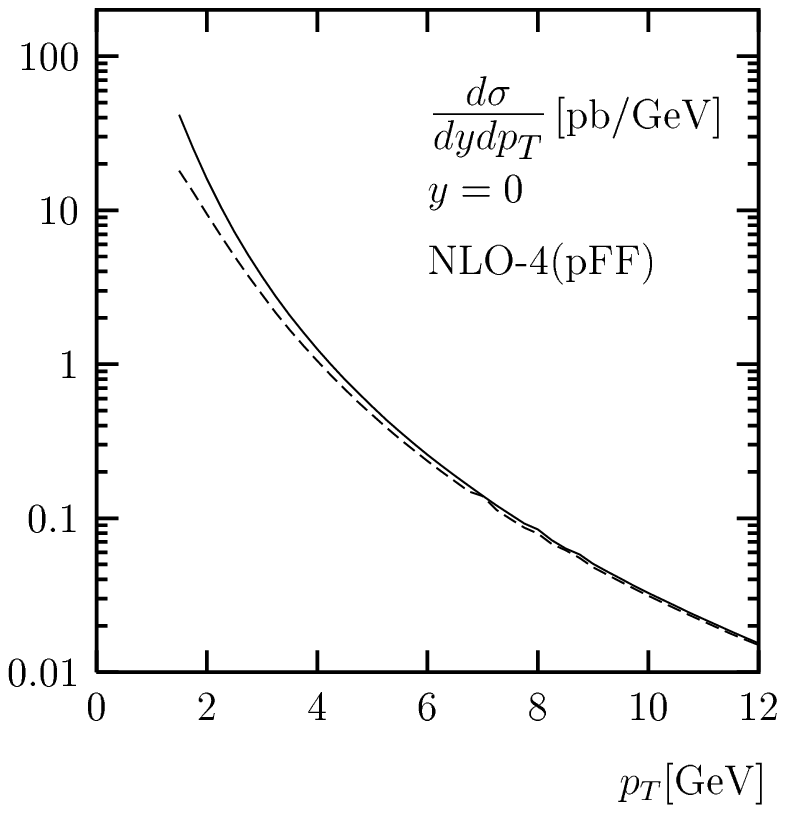,width=8cm}}
\put(40,-1){(a)}
\put(80,-1){\epsfig{file=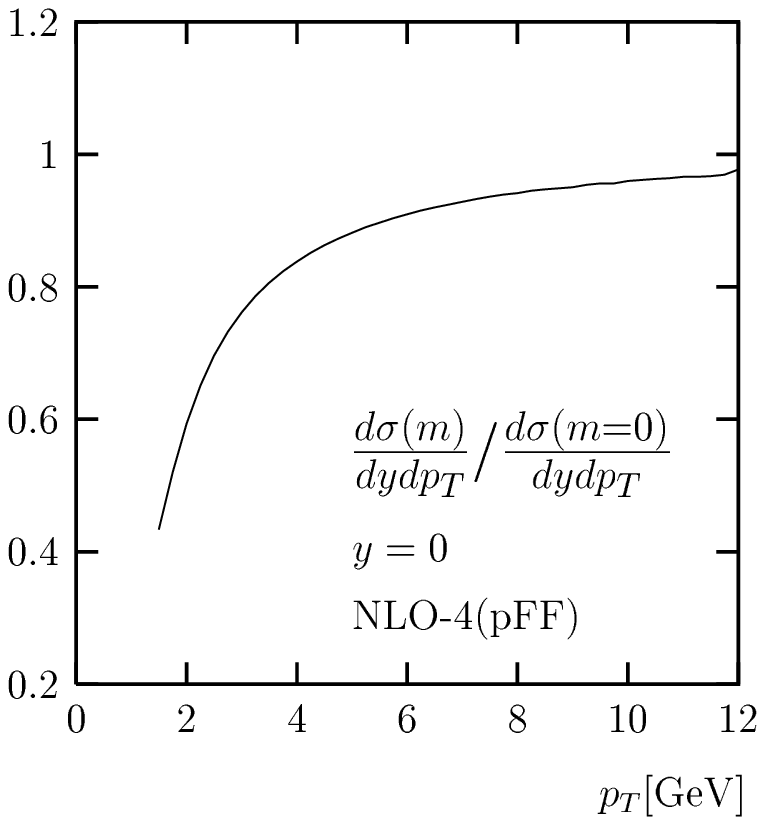,width=8cm}}
\put(120,-1){(b)}
\end{picture}
\caption{As in Fig.\ \ref{fig3}, but including $c \rightarrow D^*$
  fragmentation according to (\ref{dcc}) (perturbative FF), i.e.\
  FSI-terms resummed.}
\label{fig4}
\end{figure}

\begin{figure}[th] 
\unitlength 1mm
\begin{picture}(160,85)
\put(0,-1){\epsfig{file=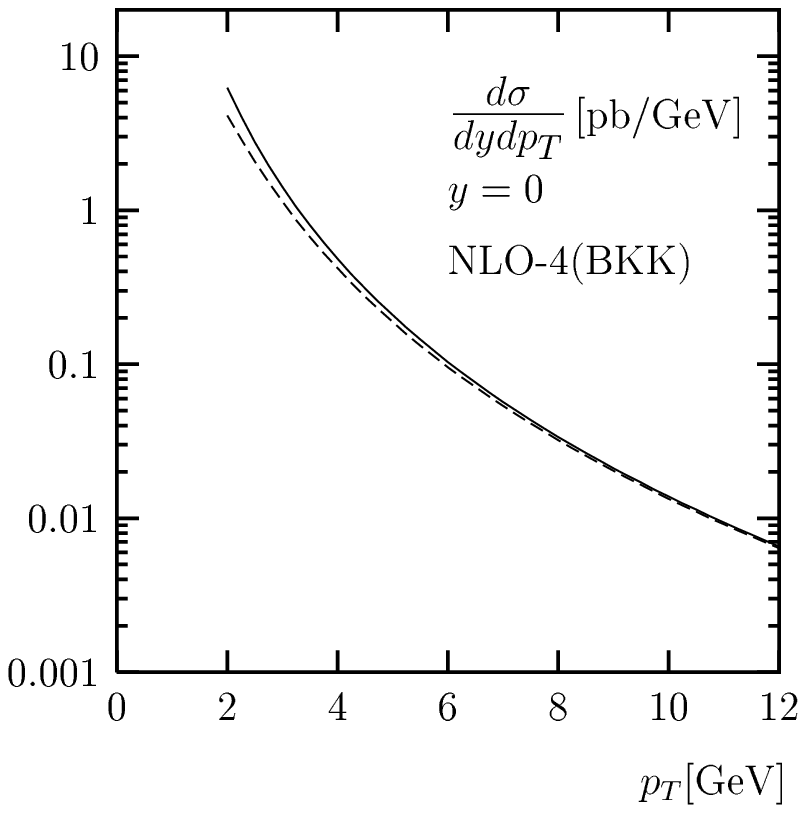,width=8cm}}
\put(40,-1){(a)}
\put(80,-1){\epsfig{file=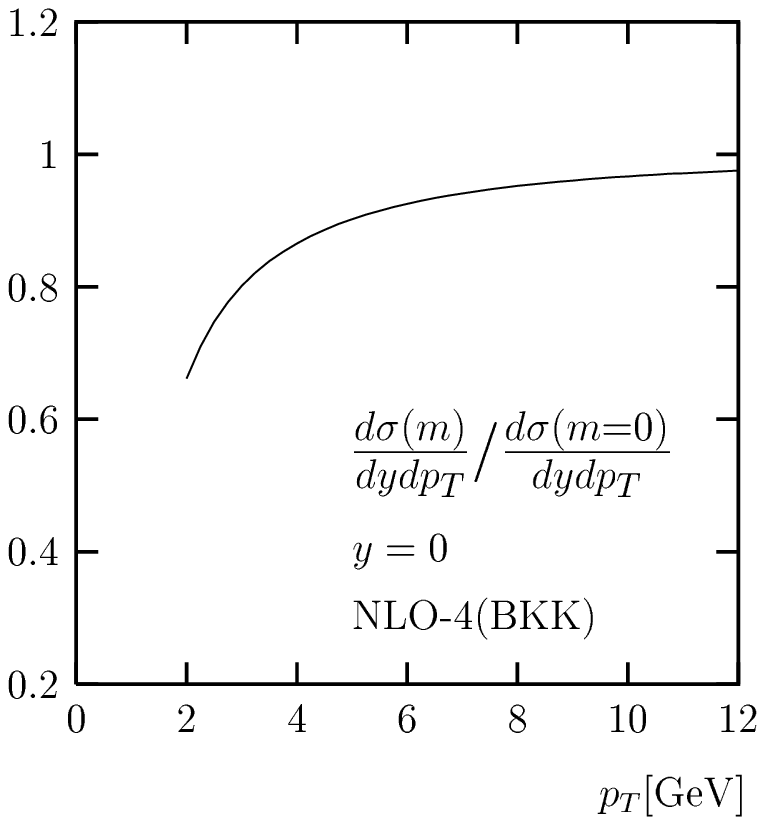,width=8cm}}
\put(120,-1){(b)}
\end{picture}
\caption{As in Fig.\ \ref{fig3}, but including $c \rightarrow D^*$
  fragmentation according to Ref.\ \cite{5} (set OPAL NLO).}
\label{fig5}
\end{figure}

It is clear that the results presented up to this point can not be
compared to experimental data as long as no FF for the transition $c
\rightarrow D^{*}$ is included. This deficiency is amended now for the
cross section $d\sigma (y=0)/dydp_T$ shown in Fig.\ \ref{fig4}a and
Fig.\ \ref{fig5}a with two different choices of the FF. The cross
sections in Fig.\ \ref{fig4}a, again for the massive and the massless
version, are the cross sections in the NLO 4-flavour scheme augmented
with the so called perturbative fragmentation function (pFF) \cite{10}.
The pFF is evolved via the Altarelli-Parisi equations to the appropriate
scale with the initial condition at scale $\mu _0$ equal to
\begin{equation}
D_c^c(x,\mu _0) = \delta (1-x) + d_c^c(x,\mu _0)
\label{dcc_init}
\end{equation}
where $d_c^c(x,\mu _0)$ is given in (\ref{dcc}) with the choice $\mu _0
= m$ as initial scale. This prescription amounts to taking into account
FSI terms not only at order $O(\alpha_s)$, but to resum them by virtue
of the Altarelli-Parisi equations.  The normalization of the cross
sections again includes the experimental branching ratio as in the
previous figures.  Such a FF can be considered as a perturbative model
for the real FF which we can use instead of a calculation based on
non-perturbative methods.  Alternatively, a FF can be extracted from
experimental data for the production of $D^{*}$ mesons obtained from
other measurements.  The latter approach will be discussed below. In
Fig.\ \ref{fig4}a the perturbative FF is applied equally to the massive
and massless cross sections. Through the Altarelli-Parisi evolution the
terms $\propto \alpha_s(\mu) \ln (M_F/\mu _0)$ are resummed, which leads
to an appreciable reduction of the cross sections at large $p_T$.
Comparing to the cross sections in Fig.\ \ref{fig3}a the reduction
amounts to a factor of more than two at $p_T=12$ GeV. In the small-$p_T$
region the influence of the fragmentation is largely reduced.  The cross
section with the pFF based on the boundary condition (\ref{dcc_init})
was used in the massless approximation for a comparison with the massive
NLO 3-flavour theory in Ref.\ \cite{9}. Comparing with Fig.\ \ref{fig2}a
we see that the two cross sections are approximately equal at $p_T
\simeq 5$ GeV and the NLO 3-flavour scheme has a much larger cross
section than the ZM 4-flavour scheme at $p_T = 12$ GeV in agreement with
the results in Fig.\ \ref{fig1}a of Ref.\ \cite{9}.  The difference at
larger $p_T$ is due to the choice of scheme and the effect of the
evolution of the FF in the ZM 4-flavour scheme.  Of course, as in the
previous figures the ZM cross section approaches the massive one for
large $p_T$. The ratio of the two cross sections with $m \neq 0$ and
$m=0$ in the NLO 4-flavour (pFF) approach is exhibited in Fig.\ 
\ref{fig4}b showing that the ratio goes to one for large $p_T$.  The
ratio is larger than $0.9$ for $p_T \geq 5.6$ GeV, i.e.\ the crossing
point occurs at a smaller $p_T$ than in the NLO 4-flavour cross section
without FF (see Fig.\ \ref{fig3}b).
\\

The approach based on perturbative fragmentation functions is, however,
not sufficient to describe the fragmentation $c \rightarrow D^{*}$. For
example, in Ref.\ \cite{19} it was shown that in order to account for
the inclusive production of $D^{*}$ mesons in $e^+e^-$ annihilation at
various center-of-mass energies an additional non-perturbative component
is needed.  This means that the fragmentation of charm quarks into
$D^{*}$ mesons cannot be calculated in perturbation theory. A
non-perturbative component, which is not known theoretically and has to
be determined from other data, is always needed. Hence, it is more
appropriate, to give up the perturbative component of the FF input
altogether and to describe the $c \rightarrow D^{*}$ transition entirely
by a non-perturbative FF, as it is done for the fragmentation of $u, d$
and $s$ quarks into light mesons.  This approach has been followed in
Ref.\ \cite{5}. There a non-perturbative FF for $c \rightarrow D^{*}$
was constructed by fitting data on $e^+ + e^- \rightarrow D^{*} +X$ at
the $Z$ mass. For the fits two data sets from ALEPH \cite{20} and OPAL
\cite{21} have been used.  Since these two data sets are not compatible
with each other, two separate fits have been performed. In the following
we shall employ the results obtained from the fit to the OPAL data
\cite{5}. LEP data are best suited for such fits since they have a
better accuracy as compared to lower energy data (see Ref.\ \cite{5} for
details). However, with LEP data there is the additional complication
that one has to separate two different sources of charmed hadrons: in
hadronic $Z$ decays charmed hadrons are expected to be produced either
directly through the hadronization of charmed quarks in the process $Z
\rightarrow c\bar{c}$ or via weak decays of $B$ hadrons produced in $Z
\rightarrow b\bar{b}$ with approximately equal rate. With the data from
Refs.\ \cite{20,21} it was possible to disentangle these two components
and to construct the FF for $c \rightarrow D^{*}$ by determining the FF
at the starting scale $\mu_0=2m$. At this initial value the charm-quark
FF was assumed to have the Peterson form, which is particularly suitable
to describe FF's that peak at large $x$. The Peterson FF depends only on
two parameters, the normalization factor $N$ and the shape parameter
$\epsilon$.  According to Ref.\ \cite{5} the fit to the OPAL data gave
in NLO the values $N=0.267$ and $\epsilon = 0.116$ resulting in a
branching ratio $BR(c \rightarrow D^{*}) = 0.238$ at $\sqrt{S} = m_Z$ in
good agreement with the average value from Refs.\ \cite{17,18}.
\\

Based on the purely non-perturbative FF of Ref.\ \cite{5} (OPAL set at
NLO) we have calculated $d\sigma /d\eta dp_T$ at $\eta =0$ for massive
and massless quarks as a function of $p_T$. $\eta $ is the
pseudo-rapidity which was used also in the analysis of the experimental
data \cite{1,2,3} and we identify the pseudo-rapidity of the $D^{*}$
with the rapidity of the charm quark.  The resulting cross sections,
denoted NLO-4(BKK), are shown for 2 GeV $ < p_T <$ 12 GeV in Fig.\ 
\ref{fig5}a.  The cross section is smaller than the one in Fig.\ 
\ref{fig4}a. This results essentially from the fact that the
non-perturbative FF peaks at a value $x<1$, whereas the perturbative FF
is dominated by the contribution proportional to $\delta(1-x)$. For a
given transverse momentum $p_T$ of the observed $D^{\ast}$, smaller $x$
in the FF probes larger $\hat{p}_T = p_T/x$ of the charm quark in the
underlying hard scattering process where the cross section is smaller.
The approach of the massive to the massless cross section is even faster
than in the previous studies.  This is seen in Fig.\ \ref{fig5}b, where
the ratio is plotted.  The ratio exceeds $0.9$ already at $p_T=4.7$ GeV.
This stronger reduction of mass corrections is also due to the different
$x$-dependence of the fragmentation functions since mass terms
proportional to $\propto m^2/p_T^2$ decrease faster with increasing
$p_T$ than the massless cross section.
\\

\begin{figure}[th] 
\unitlength 1mm
\begin{picture}(160,85)
\put(40,-1){\epsfig{file=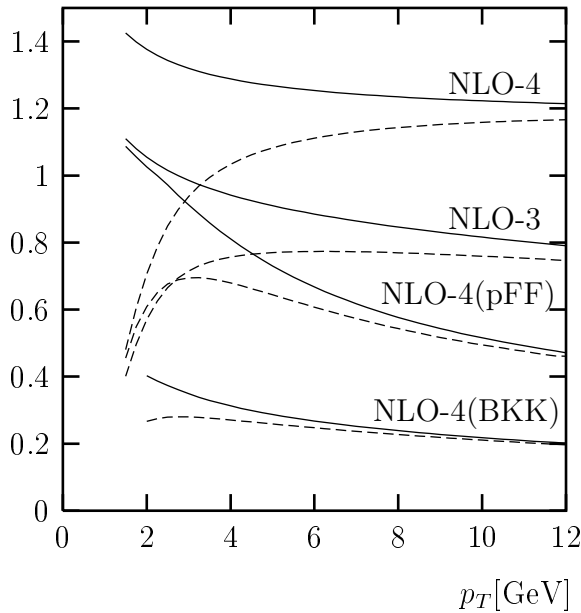,width=8cm}}
\put(100,69){NLO-4}
\put(100,51){NLO-3}
\put(91,41){NLO-4(pFF)}
\put(90,25.5){NLO-4(BKK)}
\end{picture}
\caption{$p_T$ distributions normalized to the result of the LO massless
  calculation (full curve in Fig.\ \ref{fig1}a), $d\sigma / d\sigma_{\rm
    LO}(m=0)$ at $y=0$. Full lines: massless calculations, dashed lines:
  massive calculation. From top to bottom: NLO 4-flavour scheme with
  $M_I = M_F = \sqrt{p_T^2 + m^2}$ (NLO-4); NLO 3-flavour scheme with
  $M_I = M_F = m$ (NLO-3); NLO 4-flavour scheme with $M_I = M_F =
  \sqrt{p_T^2 + m^2}$ including perturbative fragmentation (NLO-4(pFF));
  and NLO 4-flavour scheme with non-perturbative fragmentation
  (NLO-4(BKK)).}
\label{fig6}
\end{figure}

In Fig.\ \ref{fig6} we show how the NLO results in Figs.\ \ref{fig2},
\ref{fig3}, \ref{fig4} and \ref{fig5} are related to each other. For
this purpose we introduce a common normalization by dividing the results
of all four, massless and massive, cross sections by the massless LO
cross section (full curve in Fig.\ \ref{fig1}a). The full lines in Fig.\ 
\ref{fig6} represent the massless calculations for NLO-4, NLO-3,
NLO-4(pFF) and NLO-4(BKK). The dashed curves correspond to the massive
cross sections.  Compared to the LO massless cross section, the NLO
corrections in the NLO 4-flavour scheme produce an increase of the cross
section between $40\,\%$ (low $p_T$, massless) and $20\,\%$ (large
$p_T$).  The increase at low $p_T$ is compensated by mass effects and at
large $p_T$ by including the pFF so that the net effect of NLO
contributions in the range 3 GeV $< p_T < 12$ GeV is a reduction between
$30\,\%$ and $50\,\%$. Most of the reduction of the NLO 3-flavour result
compared to the NLO-4 curve originates from the different choice of the
factorization scale.  At low $p_T$, around $p_T \simeq 2$ GeV, the
massive cross sections are approximately equal to each other.  The
inclusion of a fragmentation function leads to a further reduction of
the cross section which is particularly strong for the case of the
non-perturbative FF.
\\

From this section we conclude that the corrections due the finite charm
mass are appreciable at low $p_T$ and are small, i.e.\ below $10\,\%$
for $p_T > 5$ GeV. Furthermore, the fragmentation corrections due to the
evolution of the perturbative FF are also large, but now in the
large-$p_T$ region.  For the non-perturbative FF there is also a
reduction at smaller $p_T$ due to the fact that in this case the hard
cross section is probed at considerably larger transverse momenta of the
charm quark.  The approach of the massive theory towards the massless
approximation is always very fast.  It is strongest for the theory with
a realistic, i.e.\ non-perturbative FF. For this case, at $p_T = 2$ GeV,
the neglected mass terms lead to an overestimation of the massive result
of not more than $40\,\%$.  It is also apparent, that the pure NLO
3-flavour approach, i.e.\ without any additional non-perturbative
fragmentation corrections can give a trustworthy prediction only for
rather small $p_T$, i.e.\ for $p_T < 3$ GeV. At higher $p_T$ the effect
of evolution is essential. But even at small $p_T$, we find a reduction
by a factor 2 due to the non-perturbative FF, i.e.\ even at small $p_T$
the NLO 4-flavour approach with perturbative FF does not lead to a
realistic prediction for the cross section.
\\


\section{Comparison with LEPII Data}

Experimental data for the differential cross section $d\sigma/d\eta
dp_T$ integrated over some fixed $\eta $ region come from the three LEP
collaborations ALEPH, L3 and OPAL. The ALEPH data \cite{1} represent
$d\sigma /dp_T$ integrated over $-1.5 < \eta < 1.5$ and are averaged
over the LEPII runs with $\sqrt{S}$ in the interval 183 GeV $< \sqrt{S}
< 189$ GeV. The most recent L3 data \cite{22} are integrated over $|\eta
| < 1.4$ and averaged over 183 GeV $< \sqrt{S} < 209$ GeV.  The more
recent OPAL data (second Ref.\ in \cite{3}) are integrated over $|\eta |
< 1.5$ with the luminosity averaged $\sqrt{S} = 193$ GeV (183 GeV $<
\sqrt{S} < 202$ GeV). We compare the three data sets of Refs.\ 
\cite{1,3,22} with our calculation at $\sqrt{S} = 189$ GeV.  Besides the
differing $\sqrt{S}$ regions there are the slightly different regions of
the $\eta $ integration and possibly different anti-tagging conditions
for the outgoing electrons and positrons. We disregard these differences
and put the data points of all three experiments in one plot.  We have
checked that the different $|\eta |_{max}$ chosen by L3 on one side and
ALEPH and OPAL on the other side change the cross section by at most 
$7\,\%$ which is much below the measurement errors of the data points.
Also the different values of $\sqrt{S}$ are not expected to change the
cross section by more than the corresponding experimental errors. As an
example we checked that increasing $\sqrt{S}$ from 189 to 193 GeV
increases $d\sigma/dp_T$ by 2 (3)\,\% at $p_T = 2$ (12) GeV. The
influence of different anti-tagging conditions could not be investigated
because of insufficient information in the corresponding references.
\\

For the comparison with the experimental data we need also the
contributions of the single-resolved and double-resolved channels. They
are available only in the massless approximation.  Therefore we shall
add the direct cross section as calculated in the previous section (in
the massive and massless version) and the single-resolved and
double-resolved cross sections in the massless approximation. For this
we use the codes developed in Ref.\ \cite{5} for the calculation of the
single-resolved and double-resolved cross section for the process
$\gamma \gamma \rightarrow \pi^{\pm}~X$. They were also used in Ref.\ 
\cite{23} and in the first reference of \cite{5} for the process $\gamma
p \rightarrow D^{*\pm}~X$. In all cases we shall use the BKK
fragmentation function as described in the previous section.  We choose
as factorization scales $M_I = M_F = \xi \sqrt{p_T^2 + m^2}$ with $\xi =
2$. This allows us to calculate $d\sigma/dp_T$ down to smaller $p_T$
than previously since the starting scale of the non-perturbative FF of
Ref.\ \cite{5} was put equal to $2m$.
\\

\begin{figure}[th] 
\unitlength 1mm
\begin{picture}(160,85)
\put(40,-1){\epsfig{file=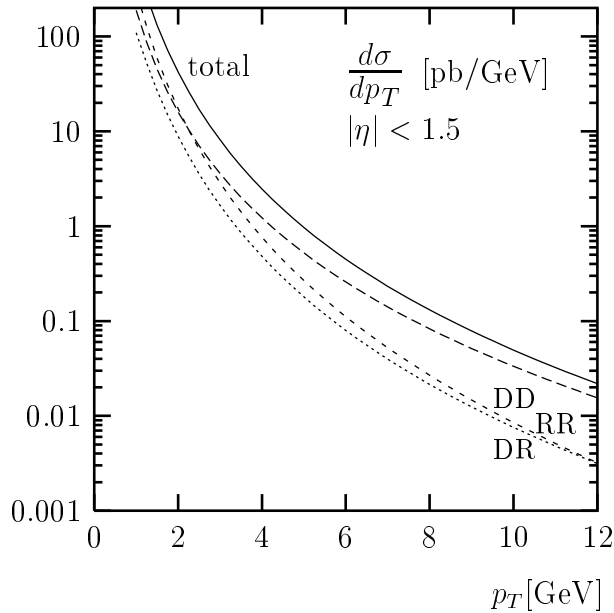,width=8cm}}
\end{picture}
\caption{Direct and resolved contributions to the $p_T$ distribution
  $d\sigma/dp_T$ after integration over $|\eta| < 1.5$ at NLO with $M_I
  = M_F = 2\sqrt{p_T^2 + m^2}$ including BKK-fragmentation. Upper
  long-dashed line: direct, dotted line: single-resolved, lower
  short-dashed line: double-resolved contributions, full line: sum of
  all contributions.}
\label{fig7}
\end{figure}

The partition of $d\sigma/d\eta dp_T$ into direct and resolved
contributions, integrated over $\eta $ in the region $-1.5 < \eta < 1.5$
is shown in Fig.\ \ref{fig7}. Here, DD, DR and RR denote the direct,
single-resolved and double-resolved cross sections in the massless
approximation and in NLO, respectively. At $p_T = 2$ GeV, these three
contributions amount to approximately $38\,\%$, $21\,\%$ and $41\,\%$ of
the total sum. At $p_T = 12$ GeV the relative contributions are
$71\,\%$, $14\,\%$ and $15\,\%$, respectively. This means that the
resolved cross sections DR and RR decrease with increasing $p_T$ much
faster than the direct component DD. The separate contributions depend
strongly on the factorization scales: for $M_I = M_F = \sqrt{p_T^2 +
  m^2}$, i.e.\ with $\xi = 1$, the direct contribution is larger by
roughly $10\,\%$ for all $p_T$, the single-resolved contribution becomes
steeper (with an increase of $38\,\%$ at $p_T = 2$ GeV and $3\,\%$ at
$p_T = 12$ GeV), whereas the double-resolved contribution is much
reduced (by $60\,\%$ at $p_T = 2$ GeV and $30\,\%$ at $p_T = 12$ GeV).
We give these numbers for completeness only, although the scale
dependence of the separate contributions is unphysical. The scale
dependence of the physically observable cross section is weak: for the
sum of all contributions these changes combine to a $13\,\%$ decrease at
$p_T = 2$ GeV and a $1\,\%$ increase at $p_T = 12$ GeV.  The
single-resolved cross section has an appreciable component originating
from the charm PDF in the photon. It is equal to $15\,\%$ at $p_T = 2$
GeV and equal to $61\,\%$ at $p_T = 12$ GeV.  The double-resolved cross
section is due entirely to the charm component in the photon. In Fig.\ 
\ref{fig7} and the following figure the photon PDF is taken from Ref.\ 
\cite{24}, which has an explicit charm contribution.
\\

\begin{figure}[ht] 
\unitlength 1mm
\begin{picture}(160,125)
\put(20,-1){\epsfig{file=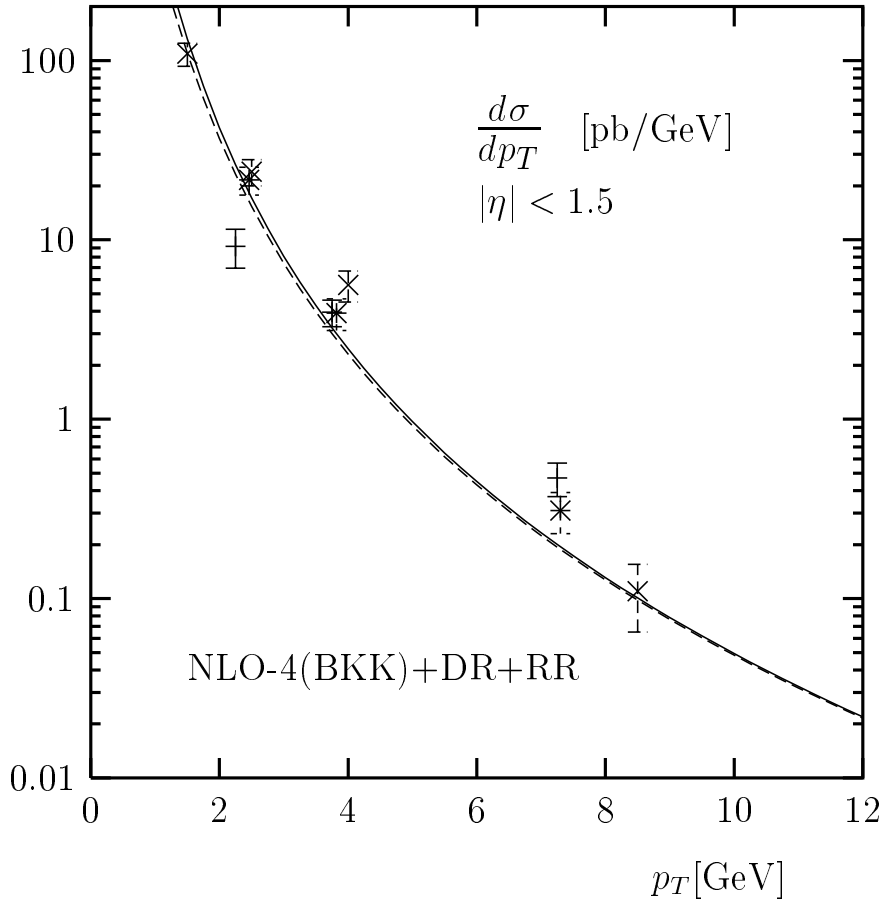,width=12cm}}
\end{picture}
\caption{$p_T$ distribution $d\sigma/dp_T$ after integration over
  $|\eta| < 1.5$ in the NLO 4-flavour scheme with $M_I = M_F =
  2\sqrt{p_T^2 + m^2}$ including BKK-fragmentation compared to LEP data
  \protect\cite{1,3,22}.  Full line: massless calculation, dashed line:
  massive calculation.  Single- and double-resolved contributions are
  included using photon PDFs of Ref.\ \protect\cite{24}.}
\label{fig8}
\end{figure}

The sum of all three contributions, DD, DR and RR, is compared to
experimental data in Fig.\ \ref{fig8}.  The full curve is the cross
section in the massless approximation as in Fig.\ \ref{fig7}.  In the
dashed curve the direct massless cross section is replaced by the direct
cross section with massive quarks, i.e.\ NLO-4(BKK) of the previous
section, except for the change of factorization scales. The resolved
components are as in Fig.\ \ref{fig7}. The experimental data points
shown at $p_T$ values between 1.5 GeV and 8.5 GeV are from ALEPH
\cite{1}, L3 \cite{22} and OPAL \cite{3}. The overall agreement between
the theoretical prediction (dashed curve) and the experimental data is
quite good although the data points in the medium $p_T$ range lie
slightly above the theoretical curve. Even if a finite charm mass
correction for the DR and RR contributions would be included, for which
we expect a reduction of the theoretical prediction by approximately
$15\,\%$ at $p_T = 2$ GeV and less at higher $p_T$, the overall
agreement for $p_T \geq 2$ GeV would hardly change.  The data from OPAL
\cite{3} and L3 \cite{22} have been compared already with the
predictions of the massless theory, which was obtained on the basis of
the work in Ref.\ \cite{5}. Compared to these results we neglected the
fragmentation of the gluon $g \rightarrow D^{\ast}$.  We checked that
these contributions are very small at small $p_T$ and amount to a
contribution of approximately $-1\,\%$ at the largest $p_T$, which is
negligible compared to the errors of the experimental data.  This is due
to the fact that the gluon enters only at NLO and that the $g
\rightarrow D^{\ast}$ FF is small compared to the dominant charm FF. A
similarly small effect results from a change of the charm mass.
Choosing $m = 1.3$ GeV increases the cross section by $1\,\%$ in the
considered $p_T$ range between 2 GeV and 12 GeV.
\\


\section{Summary and Conclusions}

In this work we compared two approaches, the massive and massless
schemes, for the calculation of inclusive charm production with 3 and 4
initial flavours.  As a first step, the cross section $d\sigma/dp_T$ is
calculated for the direct component of the $\gamma \gamma $ reaction in
NLO. We found that the massless limit of the massive cross section
differs from the massless theory with $\overline{MS}$ factorization by
finite terms which are non-singular for $m \rightarrow 0$. A first
sensible comparison of the results of the massless theory with the
massive 3-flavour approach can be performed after adding these finite
terms to the massless theory. The additional terms can be interpreted as
the order $O(\alpha_s)$ coefficient of the perturbative fragmentation
function describing final-state interactions of the transition from
massless to massive charm quarks. It is clear that a resummation of
these terms by using a perturbative fragmentation function is not
sufficient and that one needs a non-perturbative fragmentation function
(in addition to a charm distribution function in the photon). Since
fragmentation and distribution functions are usually constructed in
$\overline{MS}$ factorization, these terms must be subtracted also from
the massive theory as soon as a fragmentation function, perturbative or
non-perturbative, is taken into account.  Therefore we studied the
relation of this massive version with subtracted FSI terms with the
massless 4-flavour approach, using different assumptions concerning the
fragmentation of the $c$ quark into $D^{*}$ mesons.
\\

It turned out that the massive versions converged rather fast to their
massless limits with increasing $p_T$. The convergence was strongest
when using non-perturbative FF fitted to $e^+e^-$ annihilation data
\cite{5}. At low $p_T$, for example at $p_T = 3$ GeV, the massive cross
section is reduced by approximately $20\,\%$ as compared to the massless
approximation.  At lower energies this reduction increases and makes the
massless approximation unreliable. In addition the cross section is very
much reduced by fragmentation effects, even at low $p_T$. This reduction
increases from a factor of more than 3 at $p_T = 2$ GeV to a factor of
nearly 6 at $p_T = 12$ GeV.  Therefore, for reliable predictions one
needs a good description of the fragmentation process. In our numerical
evaluations we have taken the FF from fits to $D^{*}$ production in
$e^+e^-$ annihilation at LEPI.  To compare with recent measurements of
the inclusive $D^{*}$ cross section $d\sigma/dp_T$ in $\gamma \gamma $
reactions at LEPII we added the single- and double-resolved cross
sections, which up to now are available to us only in the massless
approximation. The agreement of the calculations and the data is quite
good down to $p_T \simeq 1.5$ GeV even if we assume that the massless
resolved contributions are reduced by a similar amount as the direct
cross section. It remains to be seen whether a fit using additional
experimental data to obtain a modified fragmentation function would lead
to a better agreement. The fact that the data at medium $p_T$ lie
somewhat above the theoretical predictions seem to indicate that a
fragmentation function which is harder than the one we used in our
analysis, could be favored.
\\

It is clear that in order to obtain results at even smaller $p_T$ the
finite charm mass corrections must be calculated also for the single-
and double-resolved cross sections. Their knowledge would also be a
prerequisite for the calculation of the total $D^{*}$ production cross
section.
\\


\section*{Acknowledgement}

We are grateful to B.A.\ Kniehl and M.\ Spira for providing us with
programs for the calculation of the single- and double-resolved
contributions.
\\


\end{document}